\documentclass[12pt]{article}
\usepackage{epsfig}
\begin{document}                   
\title{Absorption of $\phi$ mesons in near-threshold proton--nucleus reactions}
\author{E.Ya. Paryev\\
{\it Institute for Nuclear Research, Russian Academy of Sciences,}\\
{\it Moscow }}

\renewcommand{\today}{}
\maketitle
\begin{abstract}
       In the framework of the nuclear spectral function approach for incoherent primary proton--nucleon and secondary
pion--nucleon production processes we study the inclusive $\phi$ meson production in the interaction of 2.83 GeV protons
with nuclei. In particular, the A-dependences of the absolute and relative $\phi$ meson yields are investigated within the
different scenarios for its in-medium width as well as for the cross section ratio 
$\sigma_{pn \to pn{\phi}}/{\sigma_{pp \to pp{\phi}}}$.  Our model calculations take into account the acceptance
window of the ANKE facility used in a recent experiment performed at COSY.  They show that the 
pion--nucleon production channel contributes distinctly to the $\phi$ creation in heavy nuclei in the chosen kinematics and, hence,
has to be taken into consideration on close examination of the dependences of the phi meson yields on the target mass number with
the aim to get information on its width in the medium. They also demonstrate that the experimentally unknown ratio    
$\sigma_{pn \to pn{\phi}}/{\sigma_{pp \to pp{\phi}}}$ has a weak effect on the A-dependence of the relative $\phi$ meson
production cross section at incident energy of present interest, whereas it is found to be appreciably sensitive to the phi in-medium
width, which means that this relative observable can indeed be useful to help determine the above width from the direct comparison
the results of our calculations with the future data from the respective ANKE-at-COSY experiment.
\end{abstract}
                                                  
\newpage

\section*{1 Introduction}

\hspace{1.5cm} Studies of modification of the $\phi$ meson spectral function at finite baryon density through its production and
decay in pion--nucleus [1], proton--nucleus [2--6] and photon--nucleus [7--10] reactions have received considerable interest
in recent years. Because, on the one hand, the $\phi$ meson (in contrast to the $\rho$ and $\omega$ mesons) does not overlap
with other light resonances in the mass spectrum and, on the other hand, the reactions on ordinary nuclei with elementary probes
are less complicated compared to heavy--ion collisions especially in the near-threshold energy regime where a small number of
possible channels for meson production contributes, one may hope to get from these studies a more clear precursor signals for
partial restoration of chiral symmetry--the fundamental symmetry of QCD in the limit of massless quarks [11--13]--and thus to test this
prominent feature of QCD as well as to extract valuable information on the nucleon strangeness content [4, 7] and on the kaon
in-medium properties [1--4]. Furthermore, the sensitivity of such reactions to in-medium changes of hadron properties, as was
argued in [14], is expected to be comparable to that of nucleus--nucleus collisions. The modification of the $\phi$ properties
(mass and width) in matter has been studied in variuos approaches based both on the QCD sum rules [11, 15, 16] and on the
hadronic models [17--19]. All these works point at a small mass shift of the $\phi$ meson in nuclear matter
(about 1--2\% of its free mass at normal nuclear matter density $\rho_0$) and an appreciable increase (by a factor of about 
five--six [17, 18] to ten [19] at density $\rho_0$) in its in-medium total width compared to the vacuum value ($\Gamma_{\phi}=4.45$ MeV). 
Because of this in the following we will concentrate on the possible determination of the $\phi$ width in the medium at finite baryon
densities. The broadening of the $\phi$ spectral shape in the nuclear matter could be directly tested by looking at the invariant mass
spectra of its decay products. However, because of the strong final-state interactions of kaons from phi hadronic decay
 $\phi \to K^+K^-$ inside a medium, the authors of refs. [4] and [7] using, respectively, an appropriate nuclear spectral function
approach and the BUU transport model have found no measurable broadening of the resulting invariant mass distributions of the
emitted charged kaon pairs due to the $\phi$ in-medium width both from proton--nucleus and from photon--nucleus reactions. 
Since the dileptons, i.e. $e^+e^-$, ${\mu^+}{\mu^-}$ pairs from phi leptonic $\phi \to e^+e^-$, $\phi \to {\mu^+}{\mu^-}$
decay channels, experience no strong final-state interactions with nuclear matter, only the dilepton mass spectra turned out to be
sensitive to the $\phi$ renormalization at the density of ordinary nuclei when the respective low-momentum cuts to the $\phi$
three-momentum were applied [1, 4, 5]. But the measurements of these spectra are difficult due to the small cross sections for
dilepton production events [1, 4, 20]. Another indirect possibility to learn about the in-medium broadening of the $\phi$ meson
has been considered in refs. [2, 3, 6, 8--10]. As a measure for this broadening 
the A-dependence of the $\phi$ production cross section in nuclei in proton- and photon-induced reactions has been employed
in these works. The A-dependence is governed by the absorption of the $\phi$ meson flux in nuclear matter, which in turn is 
determined, in particular, by the phi in-medium width. The advantage of this method is that one can exploit the main decay channel
$\phi \to K^+K^-$, having large branching ratio ($\approx $ 50 \%), for identification of the $\phi$ meson in the phi production
experiments on cold nuclear matter, like for example, in a completed nuclear $\phi$ photoproduction experiment at SPring 8/Osaka [10]
or in proton--nucleus experiment
\footnote{It should be mentioned that the high-energy proton--nucleus experiments, in which the A-dependence of inclusive $\phi$
meson production on nuclei was studied, are reviewed in ref. [6].}
at the COSY [21], which is presently being analyzed [22]. It should be noted that the above method
has been also recently adopted to study the properties of the $\omega$ mesons [23--25] and $\Lambda (1520)$ hyperons [26] in finite
nuclei in $\gamma$- and $p$-induced reactions.     
 
        Following this method and using present models for the $\phi$ self-energy in a nuclear medium as well as an eikonal
approximation to account for the absorption of the outgoing $\phi$ meson, the authors of [3] have calculated the A-dependence
of the ratio between the total phi production cross section in heavy nucleus and a light one ($^{12}$C) in $pA$ collisions. The
calculations were performed within the local Fermi sea model at energies accessible at COSY and, in particular, at beam energy of
2.83 GeV, which was employed in the recent measurements of proton-induced $\phi$ production in nuclei at the ANKE-COSY
facility [21, 22]. They have been done also with having the poor knowledge about the free cross sections of the relevant elementary
reactions $pp \to pp{\phi}$, $pn \to pn{\phi}$ at above energies and with neglecting in part the $\phi$ creation in the two-step
processes with an intermediate pions. However, the direct comparison the results of these calculations with the future 
data from the ANKE-at-COSY experiment [21, 22] to constrain the in-medium broadening of the $\phi$ meson is troublesome, since
they were carried out for the whole available phase space of the phi and did not take into account the actual acceptance window of
the ANKE spectrometer that can detect $\phi$ mesons via decays $\phi \to K^+K^-$ only under forward polar angles in the lab system
($0^{\circ} \le \theta_{\phi} \le 8^{\circ}$) and in a limited momentum range 
($0.6~{\rm {GeV/c}} \le p_{\phi} \le 1.8~{\rm {GeV/c}}$) [22]. On the other hand, the new experimental [27, 28] and theoretical [29]
results on the $\phi$ production cross sections in free $pp$ and $pn$ interactions at initial energies of interest have been recently
obtained. In addition, the authors of ref. [2] have pointed at the secondary $\phi$ production channel ${\pi}N \to {\phi}N$ as
important one in the phi creation in heavy nuclei in near-threshold $pA$ reactions. It is clear that, in order to put a strong constraints
on the in-medium $\phi$ width from comparison the results of model calculations with the future data from the ANKE experiment [21, 22],
all of the preceding has to be properly accounted for in such calculations--the main purpose of the present investigation.

     In this paper we perform a detailed analysis of the $\phi$ production in $pA$ interactions at 2.83 GeV beam energy. We present the
detailed predictions for the A-dependences of the absolute and relative cross sections for the $\phi$ creation from these interactions
obtained in the framework of a nuclear spectral function approach [4, 30--34] for an incoherent primary proton--nucleon and secondary
pion--nucleon phi meson production processes in a different scenarios for the total $\phi$ in-medium width as well as with imposing the
ANKE spectrometer kinematical cuts on the laboratory $\phi$ momenta and production angles. In view of the future data from this
spectrometer, the predictions can be used as an important tool for extracting the valuable information on the phi in-medium properties.      

\section*{2 The model and inputs} 

\section*{2.1 Direct $\phi$ production mechanism}

\hspace{1.5cm} An incident proton can produce a $\phi$ directly in the first inelastic
$pN$ collision due to the nucleon's Fermi motion. 
Since we are interested in the near-threshold energy region, we have taken into account the following elementary processes 
which have the lowest free production threshold ($\approx$ 2.59 GeV):
\begin{equation}
p+p \to p+p+\phi,
\end{equation}
\begin{equation}
p+n \to p+n+\phi,
\end{equation}
\begin{equation}
p+n \to d+\phi.
\end{equation}
Because the phi--nucleon elastic cross section is rather small [1, 35], we will neglect the elastic ${\phi}N$ final-state interactions
in the present study. Moreover, since the $\phi$ meson pole mass in the medium is approximately not affected by medium
effects [18, 19] as well as for reason of reducing the possible uncertainty of our calculations due to the use in them the model
nucleon self-energies [36, 37], we will also ignore the medium modification of the outgoing hadron masses in the present work.
Then, taking into consideration the $\phi$ meson final-state absorption as well as assuming that the $\phi$ meson--as a narrow
resonance--is produced and propagated with its pole mass $M_{\phi}$ at small laboratory angles of our main interest and using
the results given in refs. [4, 30--33, 38], we can represent the inclusive cross section for the production on nuclei phi mesons with
the momentum ${\bf p}_{\phi}$ from the primary proton-induced reaction channels (1)--(3) as follows
\footnote{It is worth noting that for kinematical conditions of the experiment [21, 22] the contribution to the $\phi$ production
on nuclei from the phi mesons produced in main primary $pN \to pN{\phi}$ channels (1), (2) and decaying into the $K^+K^-$
pairs inside the target nucleus is found on the basis of the model developed in [4] to be negligible compared to the overal 
$\phi$ yield from primary and secondary phi creation processes calculated for these conditions within the present approach.
The latter yield corresponds to the $\phi$ decays outside the target nucleus.}
:
\begin{equation}
\frac{d\sigma_{pA\to {\phi}X}^{({\rm prim})}
({\bf p}_0)}
{d{\bf p}_{\phi}}=I_{V}[A]\left[\frac{Z}{A}\left<\frac{d\sigma_{pp\to pp{\phi}}({\bf p}_{0}^{'},M_{\phi},
{\bf p}_{\phi})}{d{\bf p}_{\phi}}\right>+\frac{N}{A}\left<\frac{d\sigma_{pn\to pn{\phi}}({\bf p}_{0}^{'},M_{\phi},
{\bf p}_{\phi})}{d{\bf p}_{\phi}}\right>\right]+
\end{equation}
$$
+I_{V}[A]\frac{N}{A}\left<\frac{d\sigma_{pn\to d{\phi}}({\bf p}_{0}^{'},M_{\phi},
{\bf p}_{\phi})}{d{\bf p}_{\phi}}\right>,
$$
where
\begin{equation}
I_{V}[A]=2{\pi}A\int\limits_{0}^{R}r_{\bot}dr_{\bot}
\int\limits_{-\sqrt{R^2-r_{\bot}^2}}^{\sqrt{R^2-r_{\bot}^2}}dz
\rho(\sqrt{r_{\bot}^2+z^2})\times
\end{equation}
$$
\times    
\exp{\left[-\sigma_{pN}^{{\rm in}}A\int\limits_{-\sqrt{R^2-r_{\bot}^2}}^{z}
\rho(\sqrt{r_{\bot}^2+x^2})dx
-\int\limits_{z}^{\sqrt{R^2-r_{\bot}^2}}\frac{dx}
{\lambda_{\phi}(\sqrt{r_{\bot}^2+x^2},M_{\phi})}\right]},
$$
\begin{equation}
\lambda_{\phi}({\bf r},M_{\phi})=\frac{p_{\phi}}{M_{\phi}
\Gamma_{{\rm tot}}({\bf r},M_{\phi})}
\end{equation}
and
\begin{equation}
\left<\frac{d\sigma_{pN\to pN{\phi}}({\bf p}_{0}^{'},M_{\phi},
{\bf p}_{\phi})}
{d{\bf p}_{\phi}}\right>=
\int\int 
P({\bf p}_t,E)d{\bf p}_tdE
\left[\frac{d\sigma_{pN\to pN{\phi}}(\sqrt{s},M_{\phi},{\bf p}_{\phi})}
{d{\bf p}_{\phi}}\right],
\end{equation}
\begin{equation}
\left<\frac{d\sigma_{pn\to d{\phi}}({\bf p}_{0}^{'},M_{\phi},
{\bf p}_{\phi})}
{d{\bf p}_{\phi}}\right>=
\int\int 
P({\bf p}_t,E)d{\bf p}_tdE
\left[\frac{d\sigma_{pn\to d{\phi}}(\sqrt{s},M_{\phi},{\bf p}_{\phi})}
{d{\bf p}_{\phi}}\right];
\end{equation}
\begin{equation}
  s=(E_{0}^{'}+E_t)^2-({\bf p}_{0}^{'}+{\bf p}_t)^2,
\end{equation}
\begin{equation}
 E_{0}^{'}=E_{0}-\frac{{\Delta{\bf p}}^2}{2M_A},
\end{equation}
\begin{equation}
 {\bf p}_{0}^{'}={\bf p}_{0}-\Delta{\bf p},
\end{equation}
\begin{equation}
 \Delta{\bf p}=\frac{E_0V_0}{p_0}\frac{{\bf p}_{0}}{|{\bf p}_{0}|},
\end{equation}
\begin{equation}
   E_t=M_A-\sqrt{(-{\bf p}_t)^2+(M_{A}-m_{N}+E)^{2}}.
\end{equation}
Here, 
$d\sigma_{pN\to pN{\phi}}(\sqrt{s},M_{\phi},{\bf p}_{\phi}) /d{\bf p}_{\phi}$ and 
$d\sigma_{pn\to d{\phi}}(\sqrt{s},M_{\phi},{\bf p}_{\phi}) /d{\bf p}_{\phi}$
are the off-shell
\footnote{The struck target nucleon is off-shell, see Eq. (13).}
differential cross sections for $\phi$ production in reactions (1), (2) and (3), respectively, 
at the $pN$ center-of-mass energy $\sqrt{s}$; $\rho({\bf r})$ and 
$P({\bf p}_t,E)$ are the density and 
nuclear spectral function normalized to unity; 
${\bf p}_t$ and $E$ are the internal momentum and removal energy of the struck target nucleon 
just before the collision; $\sigma_{pN}^{{\rm in}}$ and $\Gamma_{{\rm tot}}({\bf r},M_{\phi})$  
are the inelastic cross section
\footnote{We use $\sigma_{pN}^{{\rm in}}=30$ mb for considered projectile proton energy of 2.83 GeV [33].}
of free $pN$ interaction and total $\phi$ width in its rest frame, taken at the point ${\bf r}$ inside the nucleus
and at the pole of the resonance;
$Z$ and $N$ are the numbers of protons and neutrons in 
the target nucleus ($A=N+Z$), $M_{A}$  and $R$ are its mass and radius
\footnote{It is determined from the relation $\rho_{N}(R)=0.03\rho_{0}$ [1], where $\rho_{N}$ is the nuclear density, and is
equal to 4.0, 5.259, 6.112, 7.417, 8.737, 9.214 fm, respectively, for $^{12}$C, $^{27}$Al, $^{63}$Cu, $^{108}$Ag, $^{197}$Au, $^{238}$U 
target nuclei considered in the present work.}
; 
$m_{N}$ is the bare nucleon mass; ${\bf p}_0$ and $E_0$
are the momentum and total energy of the initial proton; $V_0$ is the nuclear optical potential that 
this proton, having the kinetic energy $T_0$ of about a few GeV, feels in the 
interior of the nucleus ($V_0 \approx 40~{\rm MeV}$).

   The first term in Eq. (4) describes the contribution to the $\phi$ meson production on nuclei from the primary $pN$ interactions
(1) and (2), whereas the second one represents the contribution to this production from the elementary process (3).

   Let us now specify the off-shell differential cross sections
$d\sigma_{pN\to pN{\phi}}(\sqrt{s},M_{\phi},{\bf p}_{\phi}) /d{\bf p}_{\phi}$ and 
$d\sigma_{pn\to d{\phi}}(\sqrt{s},M_{\phi},{\bf p}_{\phi}) /d{\bf p}_{\phi}$
for $\phi$ production in the reactions (1), (2) and (3), entering into Eqs. (4), (7), (8). Following refs. [4, 32, 39--41], we assume that these
cross sections are equivalent to the respective on-shell cross sections calculated for the off-shell kinematics of the elementary
processes  (1)--(3). In our approach the differential cross sections for $\phi$ production in the reactions (1), (2) have been
described by the three-body phase space calculations normalized to the corresponding total cross sections
$\sigma_{pN \to pN{\phi}}(\sqrt{s})$ [4, 30]: 
\begin{equation}
\frac{d\sigma_{pN\rightarrow pN{\phi}}(\sqrt{s},M_{\phi},{\bf p}_{\phi})}
{d{\bf p}_{\phi}}
={\frac{{\pi}}{4E_{\phi}}}
{\frac{\sigma_{pN\rightarrow pN{\phi}}({\sqrt{s}})}
{I_{3}(s,m_p,m_N,M_{\phi})}}
{\frac{\lambda(s_{pN},m_{p}^{2},m_{N}^{2})}{s_{pN}}},
\end{equation}
\begin{equation}
I_{3}(s,m_p,m_{N},M_{\phi})=(\frac{{\pi}}{2})^2
\int\limits_{(m_{p}+m_{N})^2}^{({\sqrt{s}}-M_{\phi})^2}
{\frac{\lambda(s_{pN},m_{p}^{2},m_{N}^{2})}{s_{pN}}}
{\frac{\lambda(s,s_{pN},M_{\phi}^{2})}{s}\,ds_{pN}},
\end{equation}
\begin{equation}
\lambda(x,y,z)=\sqrt{{\left[x-({\sqrt{y}}+{\sqrt{z}})^2\right]}{\left[x-
({\sqrt{y}}-{\sqrt{z}})^2\right]}},
\end{equation}
\begin{equation}
s_{pN}=s+M_{\phi}^{2}-2(E_{0}^{'}+E_t)E_{\phi}+
2({\bf p}_{0}^{'}+{\bf p}_t){\bf p}_{\phi}.
\end{equation}
Here, $E_{\phi}$ is the total energy of a $\phi$ meson ($E_{\phi}=\sqrt{p_{\phi}^2+M_{\phi}^2}$).

        For the free total cross section $\sigma_{pp \to pp{\phi}}(\sqrt{s})$ we have used the following parametrization:
\begin{equation}
  \sigma_{pp \to pp{\phi}}({\sqrt{s}})=10\left(1-\frac{s_{th}}{s}\right)^{1.26}\left(\frac{s_{th}}{s}\right)^{1.66}~[{\rm {{\mu}b}}],
\end{equation}
where $\sqrt{s_{th}}=2m_p+M_{\phi}$ is the threshold energy for $pp \to pp{\phi}$ reaction.
The comparison of the results of our calculations by (18) (dotted line) with the experimental data (full squares) close to the threshold
for this reaction from the installation ANKE-at-COSY [27] (three lowest data points corresponding to excess energies of 18.5, 34.5 and
75.9 MeV), from the DISTO Collaboration [42] (fourth data point at excess energy of 83 MeV) as well as from the measurement [43] at
higher energy (data point at excess energy of 1.644 GeV) is shown in Fig. 1. It is seen that our parametrization (18) fits quite well the
existing set of data for the $pp \to pp{\phi}$ reaction in the threshold region.

     For obtaining the total cross section of the $pn \to pn{\phi}$ reaction, where data are not available, we have employed the
following relation:
\begin{equation}
  \sigma_{pn \to pn{\phi}}({\sqrt{s}})=f\left(\sqrt{s}-\sqrt{s_{th}}\right)\sigma_{pp \to pp{\phi}}({\sqrt{s}}).
\end{equation}
In the literature there are [28, 29, 44, 45] a different options to choose the cross section ratio
$f=\sigma_{pn \to pn{\phi}}/{\sigma_{pp \to pp{\phi}}}$ in the near-threshold energy regime. Thus, for instance, the authors of
ref. [28] using their measurements of $\phi$ production in quasi-free $pn \to d{\phi}$ reaction close to threshold and
final-state-interaction theory [46] have found that very near threshold this ratio can be estimated as:
\begin{equation}
  f=\sigma_{pn \to pn{\phi}}/\sigma_{pp \to pp{\phi}}\approx 2.3.
\end{equation}
We will use this value in the present study. To see the sensitivity of A-dependences of the absolute and relative $\phi$
production cross sections in proton--nucleus collisions to the experimentally unknown cross section ratio $f$, we will also
adopt in our calculations instead of the above value the following dependence
\footnote{The excess energy $\epsilon$ in this dependence is measured in GeV.}
:
\begin{equation}
f(\epsilon)=\left\{
\begin{array}{llllll}
	3
	&\mbox{for $0< \epsilon \le 0.004~{\rm GeV}$}, \\
	&\\
                   6.607{\epsilon}^{0.1430}
	&\mbox{for $0.004~{\rm GeV}<\epsilon \le 0.007~{\rm GeV}$}, \\
                   &\\
                   6.882{\epsilon}^{0.1512}
                   &\mbox{for $0.007~{\rm GeV}<\epsilon \le 0.03~{\rm GeV}$}, \\
                   &\\
                   5.902{\epsilon}^{0.1074}
                   &\mbox{for $0.03~{\rm GeV}<\epsilon \le 0.08~{\rm GeV}$}, \\
                   &\\
                   5.074{\epsilon}^{0.0475}
                   &\mbox{for $0.08~{\rm GeV}<\epsilon \le 0.25~{\rm GeV}$}, \\
                   &\\
                   4.611{\epsilon}^{-0.0214}
                    &\mbox{for $\epsilon > 0.25~{\rm GeV}$},
\end{array}
\right.	
\end{equation}
which, as is seen from Fig. 2, fits fairly well (solid line) the results of recent calculations [29] (full squares) of this ratio within an
effective meson--nucleon theory. It is interesting to note that the main contribution to the $\phi$ production from the one-step
reaction channels (1), (2) at 2.83 GeV incident energy of our interest with the use in the calculations the ratio $f$ in the form (21)
and with taking into account in them the kinematical acceptance conditions of the ANKE experiment [21, 22] at COSY comes from
excess energies $\epsilon \ge 30$ MeV, where in line with Fig. 2 this ratio $>$ 4. It is thus expected that the absolute $\phi$ yield
from primary production processes (1), (2) will be enhanced when the cross section ratio
$\sigma_{pn \to pn{\phi}}/{\sigma_{pp \to pp{\phi}}}$ in the form of (21) is used (see, also, below).

      Taking into consideration the two-body kinematics of the elementary process (3), we can readily get the following
expression for the differential cross section for $\phi$ production in this process (see, also, [30, 32]):
\begin{equation}
\frac{d\sigma_{pn\rightarrow d{\phi}}(\sqrt{s},M_{\phi},{\bf p}_{\phi})}
{d{\bf p}_{\phi}}
={\frac{{\pi}}{I_2(s,m_d,M_{\phi})E_{\phi}}}
{\frac{d\sigma_{pn\rightarrow d{\phi}}({\sqrt{s}})}
{d{\bf \Omega}_{\phi}^{'}}}\times
\end{equation}
$$
\times
{\frac{1}{(\omega+E_t)}}\delta\left[\omega+E_t-\sqrt{m_d^2+({\bf Q}+{\bf p}_t)^2}\right],
$$
where
\begin{equation}
I_2(s,m_d,M_{\phi})=\frac{\pi}{2}\frac{\lambda(s,m_{d}^{2},M_{\phi}^{2})}{s},
\end{equation}
\begin{equation}
\omega=E_0^{'}-E_{\phi}, \,\,\,\,{\bf Q}={\bf p}_0^{'}-{\bf p}_{\phi}. 
\end{equation}
Here, $d\sigma_{pn\rightarrow d{\phi}}({\sqrt{s}})/d{\bf \Omega}_{\phi}^{'}$ is the differential cross section for $\phi$
production in reaction (3) in the $pn$ c.m.s. normalized to the corresponding total experimental cross section 
$\sigma_{pn \to d{\phi}}$; $m_d$ is the mass of a deuteron. The authors of ref. [28] have fitted their own experimental data on
the total cross section $\sigma_{pn \to d{\phi}}$ close to threshold by the following simple expression:
\begin{equation}
  \sigma_{pn \to d{\phi}}({\sqrt{s}})=48\sqrt{(\epsilon_d/{\rm MeV})}~[{\rm {nb}}],
\end{equation}
where $\epsilon_d=\sqrt{s}-(m_d+M_{\phi})$ is the respective c.m.s. excess energy. This expression has been used in our
calculations. In them, the angular distribution $d\sigma_{pn\rightarrow d{\phi}}({\sqrt{s}})/d{\bf \Omega}_{\phi}^{'}$
was also assumed to be isotropic [28]. 

       For the $\phi$ production calculations in the case of  $^{12}$C and $^{27}$Al, $^{63}$Cu, $^{108}$Ag, $^{197}$Au, $^{238}$U 
target nuclei reported here we have employed for the nuclear density $\rho({\bf r})$,
respectively, the harmonic oscillator and the Woods-Saxon distributions:
\begin{equation}
\rho({\bf r})={\rho}_{N}({\bf r})/A=\frac{(b/\pi)^{3/2}}{A/4}\left\{1+
\left[\frac{A-4}{6}\right]br^{2}\right\}\exp{(-br^2)},
\end{equation}
\begin{equation}
 \rho({\bf r})=\rho_{0}\left[1+
\exp{\left(\frac{r-R_{1/2}}{a}\right)}\right]^{-1}
\end{equation}
with $b=0.355~{\rm fm}^{-2}$ [4] and $R_{1/2}=3.347~{\rm fm}$ for $^{27}$Al, $R_{1/2}=4.20~{\rm fm}$ for $^{63}$Cu,
$R_{1/2}=5.505~{\rm fm}$ for $^{108}$Ag, $R_{1/2}=6.825~{\rm fm}$ for $^{197}$Au, $R_{1/2}=7.302~{\rm fm}$ for $^{238}$U,
$a=0.55~{\rm fm}$ for all nuclei [4, 47]. The nuclear spectral function $P({\bf p}_t,E)$ (which represents the
 probability to find a nucleon with momentum ${\bf p}_t$ and removal energy $E$ in the nucleus) for $^{12}$C target nucleus was taken
 from [30]. The single-particle part of this function for medium $^{27}$Al, $^{63}$Cu and heavy $^{108}$Ag, $^{197}$Au, $^{238}$U 
target nuclei was assumed to be the same as that for $^{208}$Pb [33, 34, 48]. The latter was taken from [32]. The correlated part of the
nuclear spectral function for these target nuclei was borrowed from [30]. 

   Let us concentrate now on the total $\phi$ in-medium width appearing in (6) and used in the subsequent calculations of phi meson
attenuation in $pA$ interactions.

   For this width, the calculations foresee four different scenarios: {\bf i)} no in-medium effects and, correspondingly, the scenario with the
free $\phi$ width (dotted line in Fig. 3); {\bf ii)} the same in-medium effects on the masses of a daughter kaon, antikaon $\rho$ meson and
through this on the $\phi$ decay width in a nuclear environment as well as the same collisional broadening
\footnote{Which is characterized by the collisional width of 10 MeV at saturation density $\rho_0$ [4].}
of a $\phi$ meson in this environment as those adopted before in [4] (solid line
\footnote{It shows that in this scenario the resulting total (decay+collisional) width of the $\phi$ meson grows as a function of the
density and reaches the value of around 30 MeV at the density $\rho_0$.}
in Fig. 3); {\bf iii)} the same in-medium effects on the phi decay width as those in the preceding case and the collisional broadening
of a $\phi$ meson, characterized by an enlarged phi collisional width of 25.4 MeV
\footnote{It should be pointed out that this value is close to the magnitude of the $\phi$ collisional width of 24 MeV obtained in [7]
at nuclear density $\rho_0$ and vanishing $\phi$ momentum.}
 at normal nuclear matter density to get for the total $\phi$ width at this density the value of about 45 MeV predicted in ref. [19]
(dashed line in Fig. 3) and {\bf iv)} the same in-medium effects on the phi decay width as those in the previous cases
and the collisional broadening of a $\phi$ meson, determined by an increased phi collisional width of 40.4 MeV
at nuclear density $\rho_0$ to gain for the $\phi$ total width at this density the value of the order of 60 MeV, which is twice the
total width at $\rho_0$, used in the second scenario (dot-dashed line in Fig. 3). The parametrizations of [4] are taken here to calculate
the density dependences of the $\phi$ decay and collisional widths. It should be noted that the adoption of the phi in-medium widths
depicted in Fig. 3 is justified not only for small [4, 19], but also for high $\phi$ momenta falling in the range, at least, up to 2 GeV/c due
to a relatively weak dependence of the imaginary part, $Im \Pi_{\phi}$, of the $\phi$ self-energy
\footnote{Let us remind that this part is tied to the phi total width in its rest frame by the
relation: $\Gamma_{\rm {tot}}=-Im \Pi_{\phi}/M_{\phi}$.}
 in nuclear matter on its momentum in this range found in [8]. We are interested in the $\phi$ momentum region mentioned above
in view of our intention to account for in our calculations of the phi production in proton--nucleus reactions the acceptance window
of the ANKE spectrometer belonging namely to this region.

   Let us consider now the two-step $\phi$ production mechanism.

\section*{2.2 Two-step $\phi$ production mechanism}

\hspace{1.5cm} At the bombarding energy of our interest (2.83 GeV) the following two-step $\phi$ production process with a pion
\footnote{Which is assumed to be on-shell.}
in an intermediate state may contribute to the phi production in $pA$ interactions:
\begin{equation}
p+N_1 \to \pi+X,
\end{equation}
\begin{equation}
\pi+N_2 \to \phi+N,
\end{equation}
provided that the latter subprocess is allowed energetically
\footnote{We remind that the free threshold energy for this subprocess amounts to 1.43 GeV.}
.
Taking into account the phi final-state absorption and ignoring the influence 
\footnote{In line with the assumption about the absence of such influence on the final hadron masses in primary proton-induced
reaction channels (1)--(3).}
of the nuclear environment
on the outgoing hadron masses in the $\phi$ production channel (29) as well as using the results given in [33, 38], we get the
following expression for the phi production cross section for $pA$ reactions at small laboratory angles of interest from this
channel:
\begin{equation}
\frac{d\sigma_{pA\to {\phi}X}^{({\rm sec})}
({\bf p}_0)}
{d{\bf p}_{\phi}}=\frac{I_{V}^{({\rm sec})}[A]}{I_{V}^{'}[A]}
\sum_{\pi=\pi^+,\pi^0,\pi^-}\int \limits_{4\pi}d{\bf \Omega}_{\pi}
\int \limits_{p_{\pi}^{{\rm abs}}}^{p_{\pi}^{{\rm lim}}
(\vartheta_{\pi})}p_{\pi}^{2}
dp_{\pi}
\frac{d\sigma_{pA\to {\pi}X}^{({\rm prim})}({\bf p}_0)}{d{\bf p}_{\pi}}\times
\end{equation}
$$
\times
\left[\frac{Z}{A}\left<\frac{d\sigma_{{\pi}p\to{\phi}N}({\bf p}_{\pi},
{\bf p}_{\phi})}{d{\bf p}_{\phi}}\right>+\frac{N}{A}\left<\frac{d\sigma_{{\pi}n\to{\phi}N}({\bf p}_{\pi},
{\bf p}_{\phi})}{d{\bf p}_{\phi}}\right>\right],
$$
where
\begin{equation}
I_{V}^{({\rm sec})}[A]=2{\pi}A^2\int\limits_{0}^{R}r_{\bot}dr_{\bot}
\int\limits_{-\sqrt{R^2-r_{\bot}^2}}^{\sqrt{R^2-r_{\bot}^2}}dz
\rho(\sqrt{r_{\bot}^2+z^2})
\int\limits_{0}^{\sqrt{R^2-r_{\bot}^2}-z}dl
\rho(\sqrt{r_{\bot}^2+(z+l)^2})
\times
\end{equation}
$$
\times    
\exp{\left[-\sigma_{pN}^{{\rm in}}A\int\limits_{-\sqrt{R^2-r_{\bot}^2}}^{z}
\rho(\sqrt{r_{\bot}^2+x^2})dx
-\sigma_{{\pi}N}^{{\rm tot}}A\int\limits_{z}^{z+l}
\rho(\sqrt{r_{\bot}^2+x^2})dx\right]}
\times
$$
$$
\times
\exp{\left[-\int\limits_{z+l}^{\sqrt{R^2-r_{\bot}^2}}\frac{dx}
{\lambda_{\phi}(\sqrt{r_{\bot}^2+x^2},M_{\phi})}\right]},
$$
\begin{equation}
I_{V}^{'}[A]=2{\pi}A\int\limits_{0}^{R}r_{\bot}dr_{\bot}
\int\limits_{-\sqrt{R^2-r_{\bot}^2}}^{\sqrt{R^2-r_{\bot}^2}}dz
\rho(\sqrt{r_{\bot}^2+z^2})\times
\end{equation}
$$
\times    
\exp{\left[-\sigma_{pN}^{{\rm in}}A\int\limits_{-\sqrt{R^2-r_{\bot}^2}}^{z}
\rho(\sqrt{r_{\bot}^2+x^2})dx
-\sigma_{{\pi}N}^{{\rm tot}}A\int\limits_{z}^{\sqrt{R^2-r_{\bot}^2}}
\rho(\sqrt{r_{\bot}^2+x^2})dx\right]},
$$
\begin{equation}
\left<\frac{d\sigma_{{\pi}N\to {\phi}N}({\bf p}_{\pi},
{\bf p}_{\phi})}
{d{\bf p}_{\phi}}\right>=
\int\int 
P({\bf p}_t,E)d{\bf p}_tdE
\left[\frac{d\sigma_{{\pi}N\to {\phi}N}(\sqrt{s_1},{\bf p}_{\phi})}
{d{\bf p}_{\phi}}\right];
\end{equation}
\begin{equation}
  s_1=(E_{\pi}+E_{t})^2-(p_{\pi}{\bf \Omega_{0}}+{\bf p}_{t})^2,
\end{equation}
\begin{equation}
 p_{\pi}^{{\rm lim}}(\vartheta_{\pi}) =
\frac{{\beta}_{A}p_{0}\cos{\vartheta_{\pi}}+
 (E_{0}+M_A)\sqrt{{\beta}_{A}^2-4m_{\pi}^{2}(s_{A}+
p_{0}^{2}\sin^{2}{\vartheta_{\pi}})}}{2(s_{A}+
p_{0}^{2}\sin^{2}{\vartheta_{\pi}})},
\end{equation}
\begin{equation}
 {\beta}_A=s_{A}+m_{\pi}^{2}-M_{A+1}^{2},\,\,s_A=(E_{0}+M_A)^2-p_{0}^{2},
\end{equation} 
\begin{equation}                                 
\cos{\vartheta_{\pi}}={\bf \Omega}_0{\bf \Omega}_{\pi},\,\,\,\,
{\bf \Omega}_{0}={\bf p}_{0}/p_{0},\,\,\,\,{\bf \Omega}_{\pi}={\bf p}_{\pi}/p_{\pi}.
\end{equation}
Here, $d\sigma_{pA\to {\pi}X}^{({\rm prim})}({\bf p}_0)/d{\bf p}_{\pi}$ are the
inclusive differential cross sections for pion production on nuclei at small laboratory angles and for high momenta from
the primary proton-induced reaction channel (28); $d\sigma_{{\pi}N\to 
{\phi}N}(\sqrt{s_1},{\bf p}_{\phi})/d{\bf p}_{\phi}$ is
the free inclusive differential cross section for $\phi$ production via the subprocess (29) calculated for the off-shell kinematics of
this subprocess at the ${\pi}N$ center-of-mass energy $\sqrt{s_1}$;
$\sigma_{\pi N}^{{\rm tot}}$ is the total cross section of the free $\pi N$ interaction
\footnote{We use in the following calculations $\sigma_{\pi N}^{{\rm tot}}=35$ mb for all pion momenta [33].}
;
${\bf p}_{\pi}$ and $E_{\pi}$ are the momentum and total energy of a pion;
$p_{\pi}^{{\rm abs}}$ is the absolute threshold momentum for phi production on the residual nucleus by an intermediate pion;
$p_{\pi}^{{\rm lim}}(\vartheta_{\pi})$ is the kinematical limit for pion production
at the lab angle $\vartheta_{\pi}$ from proton-nucleus collisions. The quantity $\lambda_{\phi}$ is defined above by Eq. (6).

    The expression for the differential cross section for $\phi$ production in the elementary process (29) has the form similar to
that (22) for proton-induced reaction (3):
\begin{equation}
\frac{d\sigma_{{\pi}N\rightarrow {\phi}N}(\sqrt{s_1},{\bf p}_{\phi})}
{d{\bf p}_{\phi}}
={\frac{{\pi}}{I_2(s_1,m_N,M_{\phi})E_{\phi}}}
{\frac{d\sigma_{{\pi}N\rightarrow {\phi}N}({\sqrt{s_1}})}
{d{\bf \Omega}_{\phi}^{*}}}\times
\end{equation}
$$
\times
{\frac{1}{(\omega_1+E_t)}}\delta\left[\omega_1+E_t-\sqrt{m_N^2+({\bf Q}_1+{\bf p}_t)^2}\right],
$$
where
\begin{equation}
\omega_1=E_{\pi}-E_{\phi}, \,\,\,\,{\bf Q}_1={\bf p}_{\pi}-{\bf p}_{\phi} 
\end{equation}
and the quantity $I_2$ is defined above by Eq. (23). 
In Eq. (38), $d\sigma_{{\pi}N\rightarrow {\phi}N}({\sqrt{s_1}})/d{\bf \Omega}_{\phi}^{*}$ is the 
differential cross section for $\phi$ production in reaction (29) in the ${\pi}N$ c.m.s., which is assumed to be isotropic [1, 49] in
our calculations of phi creation in $pA$ collisions from this reaction:
\begin{equation}
\frac{d\sigma_{{\pi}N\rightarrow {\phi}N}({\sqrt{s_1}})}
{d{\bf \Omega}_{\phi}^{*}}=\frac{\sigma_{{\pi}N\rightarrow {\phi}N}({\sqrt{s_1}})}{4\pi}.
\end{equation}
Here, $\sigma_{{\pi}N\rightarrow {\phi}N}({\sqrt{s_1}})$ is the total cross section of the elementary process ${\pi}N \to {\phi}N$.
The elementary $\phi$ production reactions  ${\pi}^+n \to {\phi}p$, ${\pi}^0p \to {\phi}p$, ${\pi}^0n \to {\phi}n$ and
${\pi}^-p \to {\phi}n$ have been included in our calculations of the $\phi$ production on nuclei.
The isospin considerations show that the following relations among the total cross sections of these reactions exist [50]:
\begin{equation}
\sigma_{{\pi}^{+}n \to {\phi}p}=\sigma_{{\pi}^{-}p \to {\phi}n},
\end{equation}
\begin{equation}
\sigma_{{\pi}^{0}p \to {\phi}p}=\sigma_{{\pi}^{0}n \to {\phi}n}=\frac{1}{2}\sigma_{{\pi}^{-}p \to {\phi}n}.
\end{equation}
For the free total cross section
$\sigma_{{\pi}^{-}p \to {\phi}n}$ we have used the following parametrization suggested in [51]:
\begin{equation}
\sigma_{{\pi}^-p \to {\phi}n}({\sqrt{s_1}})=\left\{
\begin{array}{ll}
	0.47\left(\sqrt{s_1}-\sqrt{s_0}\right)~[{\rm mb}]
	&\mbox{for $\sqrt{s_0}<\sqrt{s_1}< 2.05~{\rm GeV}$}, \\
	&\\
                   23.7/s_1^{4.4}~[{\rm mb}]
	&\mbox{for $\sqrt{s_1}\ge 2.05~{\rm GeV}$}, 
\end{array}
\right.	
\end{equation}
where $\sqrt{s_0}=m_n+M_{\phi}$ is the threshold energy.

  It should be noted that for the cross section $\sigma_{{\pi}^{-}p \to {\phi}n}$ we have also adopted another parametrization:
\begin{equation}
\sigma_{{\pi}^-p \to {\phi}n}({\sqrt{s_1}})=\left\{
\begin{array}{ll}
	0.101\left(\sqrt{s_1}-\sqrt{s_0}\right)^{0.466}~[{\rm mb}]
	&\mbox{for $\sqrt{s_0}<\sqrt{s_1}< 2.08~{\rm GeV}$}, \\
	&\\
                   23.7/s_1^{4.4}~[{\rm mb}]
	&\mbox{for $\sqrt{s_1}\ge 2.08~{\rm GeV}$}, 
\end{array}
\right.	
\end{equation}
which combines the relevant parametrization from [50] in the "low" energy region with that
given by Eq. (43) in the "high"
energy region. The parametrization (44) is close to the results from the boson-exchange model [52] near the threshold,
where the data are not available, and is considerably larger here than the one given above 
by Eq. (43). However, its use instead of (43),
as showed our calculations, leads to increase of the relative cross sections for $\phi$ production in $pA$ collisions of main interest
only by about two percents at beam energy of 2.83 GeV. Therefore, this gives confidence to us that the expression (43) is reliable enough
to describe the $\phi$ production in these collisions through the ${\pi}N \to {\phi}N$ reactions. We will employ this expression
throughout our calculations.

   Another a very important ingredients for the calculation of the phi production cross section in proton--nucleus reactions
from pion-induced reaction channel (29)--the high momentum parts of the differential cross sections for pion production on
nuclei at small lab angles from the primary process (28)--for $^{63}$Cu target nucleus were taken from [33]. For $\phi$
production calculations in the case of $^{12}$C and $^{27}$Al, $^{108}$Ag, $^{197}$Au, $^{238}$U target nuclei presented
below we have supposed [34] that the ratio of the differential cross section for pion creation on $^{12}$C and on these nuclei
from the primary process (28) to the effective number
\footnote{Which is given by Eq. (32).}
of nucleons participating in it is the same as that for $^{9}$Be and $^{63}$Cu adjusted for the kinematics relating, respectively, to
$^{12}$C and $^{27}$Al, $^{108}$Ag, $^{197}$Au, $^{238}$U.   

  Now, let us proceed to the discussion of the results of our calculations for phi production in $pA$ interactions in the
framework of the model outlined above.

\section*{3 Results and discussion}

\hspace{1.5cm} At first, we consider the A-dependences of the absolute $\phi$ production cross sections from the one-step and 
two-step
phi creation mechanisms in $pA$ collisions calculated on the basis of Eqs. (4), (30) for proton kinetic energy of 2.83 GeV in two
scenarios for the total $\phi$ in-medium width as well as with imposing the ANKE spectrometer kinematical cuts on the laboratory
phi momenta and production angles. These dependences are depicted in Figs. 4 and 5. Looking at these figures, one can see that the
one-step $\phi$ production process $pn \to d{\phi}$, with the kinematic cuts imposed, plays a minor role for all considered target
nuclei, whereas the role of secondary pion-induced reaction channel ${\pi}N \to {\phi}N$ is not negligible compared to that of 
primary $pN \to pN{\phi}$ processes in the chosen kinematics for nuclei larger than $^{27}$Al especially in the case when
the cross section ratio $\sigma_{pn \to pn{\phi}}/{\sigma_{pp \to pp{\phi}}}$ was taken to be equal to 2.3
\footnote{For small nuclei $^{12}$C and $^{27}$Al the primary processes (1), (2), as can be seen from Figs. 4 and 5, clearly dominate
the $\phi$ production for both options (20), (21) for the ratio $\sigma_{pn \to pn{\phi}}/{\sigma_{pp \to pp{\phi}}}$. It should be
noted that this fact is in line with the conclusion deduced in [53] about the dominant role of the direct $\phi$ production mechanism
in the above threshold phi creation in $p^{12}$C interactions.}
.  
This means that the channel ${\pi}N \to {\phi}N$ has to be taken into consideration on close examination of the A-dependence
of the relative $\phi$ meson production cross section in proton--nucleus reactions at energies just above threshold with the aim
of extracting of the information on the $\phi$ width in nuclear medium. Comparing the curves, corresponding to the calculations
of the $\phi$ yield from the primary $pN \to pN{\phi}$ processes (1), (2) for the two employed options for the ratio
$\sigma_{pn \to pn{\phi}}/{\sigma_{pp \to pp{\phi}}}$, one can also see that the use for this ratio the
excess-energy-dependent form (21) leads to increase of the phi yield from these processes by about a factor of 2 compared to that
obtained in the case when this ratio amounts to 2.3 for both considered scenarios for the $\phi$ width. This observation can be
useful to help constrain the above ratio from the study of the $\phi$ meson production in near-threshold proton--nucleus collisions.
Comparing Figs. 4 and 5, we see yet that the results of calculations presented in Fig. 5 are reduced (by a factor of about two) for
heavy nuclei compared to those shown in Fig. 4 due to the increased phi width in the medium and, hence, the $\phi$ absorption here,
in the case of Fig. 5.

    We, therefore, come to the conclusion that the in-medium properties of the phi should be in principle observable through the
target mass dependence of the absolute $\phi$ production cross section in $pA$ reactions at above threshold beam energies.

    However, the authors of ref. [3] have suggested to use as a measure for the $\phi$ meson width in nuclei the following relative
observable--the double ratio: $R(^{A}X)/R(^{12}{\rm C})=
(\sigma_{pA \to {\phi}X}/A)/( \sigma_{p^{12}{\rm C} \to {\phi}X}/12)$,
i.e. the ratio of the nuclear total phi production cross section from $pA$ reactions divided by A to the same quantity on $^{12}$C.
But instead of this ratio, we consider the following analogous ratio
$R(^{A}X)/R(^{12}{\rm C})=\\
({\tilde \sigma}_{pA \to {\phi}X}/A)/({\tilde \sigma}_{p^{12}{\rm C} \to {\phi}X}/12)$, where
${\tilde \sigma}_{pA \to {\phi}X}$ is the nuclear cross section for $\phi$ production from proton--nucleus collisions in the
ANKE acceptance window: $0.6~{\rm {GeV/c}} \le p_{\phi} \le 1.8~{\rm {GeV/c}}$ and $0^{\circ} \le \theta_{\phi} \le 8^{\circ}$.
It is clear that our predictions for the latter ratio could be directly compared with the future data from the ANKE experiment [21, 22]
to extract the definite information on the $\phi$ width in the nuclear matter.

      Figure 6 shows the ratio
$R(^{A}X)/R(^{12}{\rm C})$=$({\tilde \sigma}_{pA \to {\phi}X}/A)/({\tilde \sigma}_{p^{12}{\rm C} \to {\phi}X}/12)$
as a function of the nuclear mass number A calculated on the basis of Eqs. (4), (30) for the one-step and one- plus two-step
$\phi$ creation mechanisms (corresponding lines with a shorthand symbols "prim" and "prim+sec" by them) for the projectile
energy of 2.83 GeV as well as for the cross section ratio $\sigma_{pn \to pn{\phi}}/{\sigma_{pp \to pp{\phi}}}=2.3$ and within
the first two scenarios for the $\phi$ in-medium width: {\bf i)} free phi width (dotted lines), 
{\bf ii)} in-medium phi width shown by solid curve
in Fig. 3 (solid lines). The same as in Fig. 6, but calculated for the cross section ratio 
$\sigma_{pn \to pn{\phi}}/{\sigma_{pp \to pp{\phi}}}$ in the excess-energy-dependent form (21) is given in Fig. 7.
It can be seen that there are clear differences between the results obtained by using, on the one hand, different $\phi$ in-medium
widths under consideration and the same assumptions concerning the phi production mechanism (between corresponding dotted
and solid lines), and, on the other hand, different suppositions about the $\phi$ creation mechanism and the same phi widths in the
medium (between dotted lines, and between solid lines) for both considered choices for the quantity
$\sigma_{pn \to pn{\phi}}/{\sigma_{pp \to pp{\phi}}}$. We may see, for example, that for heavy nuclei, where the $\phi$
absorption is enhanced, the calculated ratio $R(^{A}X)/R(^{12}{\rm C})$ can be of the order of 0.29 and 0.43 for the direct $\phi$
production mechanism as well as 0.35 and 0.55 for the direct plus two-step phi creation mechanisms in the cases when the absorption
of phi mesons in nuclear matter was determined, respectively, by their in-medium width shown by solid curve in Fig. 3 and by their free
width. Therefore, we can conclude that the observation of the A-dependence, like that just considered, can serve as an important
tool to determine the $\phi$ width in nuclei and in the analysis of the observed dependence it is needed to account for the secondary
pion-induced phi production processes.

      In Fig. 8 we show together the results of our calculations, given before separately in Figs. 6, 7, for the A-dependence of present
interest for the primary (1)--(3) plus secondary (29) $\phi$ production processes obtained for bombarding energy of 2.83 GeV by
employing in them both the free phi mesons width (dotted lines) and their in-medium width shown by solid curve in Fig. 3 (solid lines)
as well as the two options
\footnote{Indicated by the respective symbols by the lines.}
for the ratio $\sigma_{pn \to pn{\phi}}/{\sigma_{pp \to pp{\phi}}}$ to see more clearly the sensitivity of the calculated A-dependence
to this ratio. One can see that the differences between the calculations for the same $\phi$ widths with adopting for the quantity
$\sigma_{pn \to pn{\phi}}/{\sigma_{pp \to pp{\phi}}}$ two options (20), (21) are insignificant (within a several percents), which means
that this experimentally unknown quantity has
\footnote{Contrary to our previous findings of Figs. 4 and 5 where its influence on the absolute $\phi$ production cross sections was
found to be essential.}
a weak effect on the relative observable under consideration. Therefore, this observable can indeed be useful to help determine the phi
width in the medium.

        Finally, in Fig. 9 we show the predictions of our model for the double ratio $R(^{A}X)/R(^{12}{\rm C})$ of interest obtained for
initial energy of 2.83 GeV by considering primary proton--nucleon (1)--(3) and secondary pion--nucleon (29) $\phi$ production
processes as well as by using for the cross section ratio $\sigma_{pn \to pn{\phi}}/{\sigma_{pp \to pp{\phi}}}$
the excess-energy-dependent form (21) and for the $\phi$ in-medium width those shown by dashed and dot-dashed curves in Fig. 3
(respectively, dashed and dot-dashed lines in Fig. 9) in comparison to the previously performed calculations with employing for this
width the free one (dotted line in Fig. 9) and that depicted by solid curve in Fig. 3 (solid line in Fig. 9). We observe in this figure the
visible differences between all calculations corresponding to different scenarios for the $\phi$ width in nuclei, which means that the
future data from the ANKE-at-COSY experiment [21, 22] should help, as one may hope, to distinguish between these scenarios.

    Thus, consistent with previous findings of [3], our present results demonstrate that the measurements of the A-dependence of
the relative cross section for phi production in $pA$ reactions in the considered kinematics and at above threshold beam energies 
will allow indeed to shed light on the $\phi$ in-medium properties. They show also that to extract the definite information on the 
$\phi$ width in nuclear matter from the analysis of the measured such dependence it is needed to take into account in this analysis
the secondary pion-induced phi production processes.

\section*{4 Summary}

\hspace{1.5cm} In this paper we have calculated the A-dependences of the absolute and relative cross sections for $\phi$ production
from $pA$ reactions at 2.83 GeV beam energy by considering incoherent primary proton--nucleon and secondary pion--nucleon
phi production processes in the framework of a nuclear spectral function approach [4, 30--34], which takes properly into account
the struck target nucleon momentum and removal energy distribution, novel elementary cross sections for proton--nucleon 
reaction channel close to threshold as well as different scenarios for the total $\phi$ width in the medium, for the cross section
ratio $\sigma_{pn \to pn{\phi}}/{\sigma_{pp \to pp{\phi}}}$ and the ANKE spectrometer kinematical cuts on the laboratory phi
momenta and production angles. It was found that the secondary pion-induced reaction channel ${\pi}N \to {\phi}N$ contributes
distinctly to the $\phi$ production in heavy nuclei in the chosen kinematics and, hence, it has to be taken into consideration on close
examination of the A-dependences of the $\phi$ meson production cross sections in $pA$ interactions with the aim to get information
on the phi width in the nuclear matter. It was also shown that the experimentally unknown cross section ratio
$\sigma_{pn \to pn{\phi}}/{\sigma_{pp \to pp{\phi}}}$ has, contrary to our findings for the absolute phi production cross sections,
a weak effect on the A-dependence of the relative $\phi$ meson production cross section from $pA$ collisions in the considered
kinematics and at incident energy of interest, whereas it was found to be appreciably sensitive to the absorption of the $\phi$ in the
surrounding nuclear matter in its way out of the nucleus, which is governed in turn by its in-medium width. This gives a nice 
opportunity to obtain the definite information on the $\phi$ width in the nuclear medium at finite baryon densities from the direct
comparison the results of our present calculations for this relative observable with the future data from the recently performed
ANKE-at-COSY experiment [21, 22].

            The author acknowledges valuable discussions with M. Hartmann, Yu. Kiselev, V. Koptev, A. Sibirtsev, 
H. Str$\ddot{\rm o}$her. This work is partly supported by the Russian Fund for Basic Research, Grant No.07-02-91565.

\newpage
\begin{figure}[h!]
\centerline{\epsfig{file=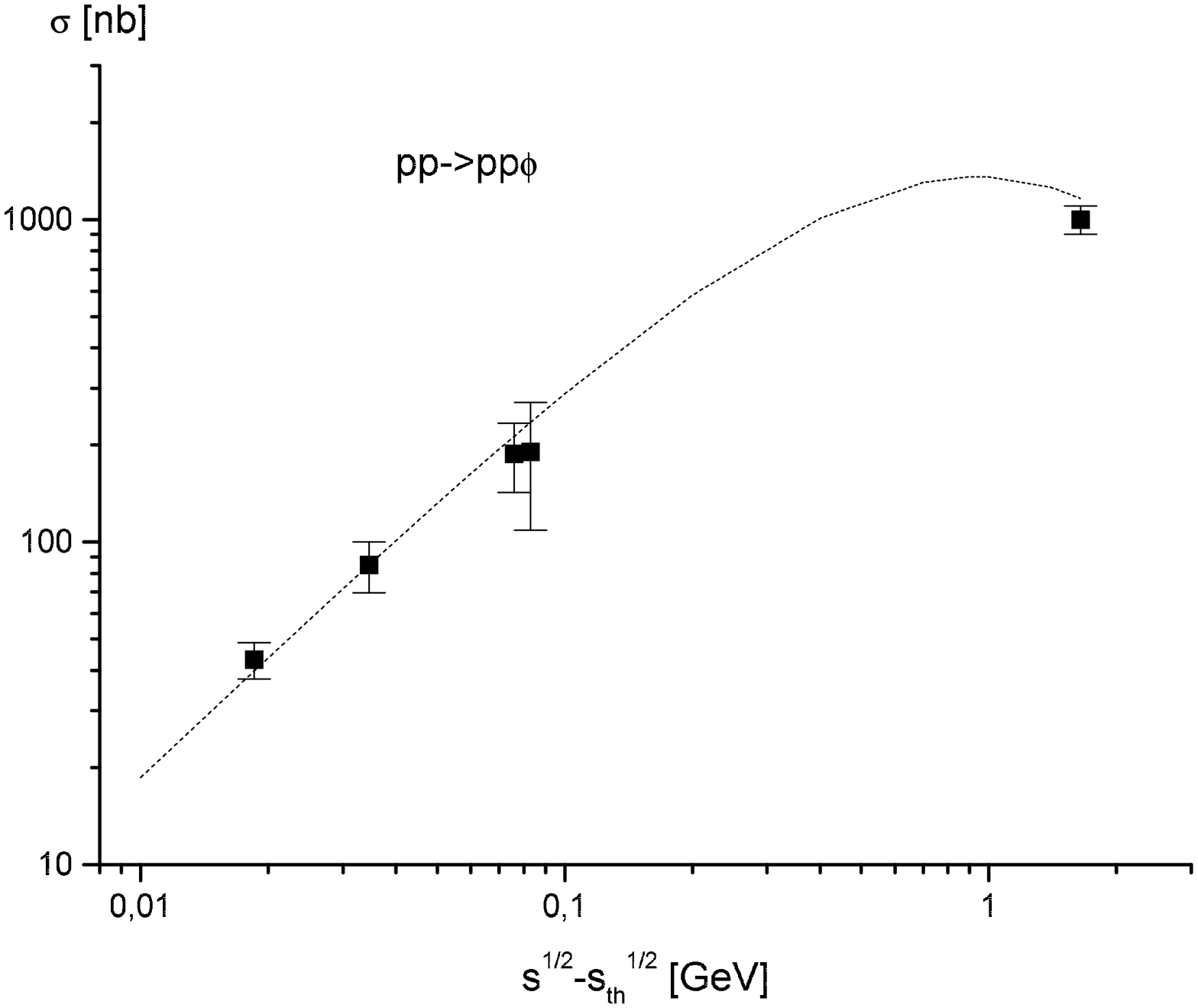,width=.88\textwidth,angle=270,silent=,
clip=}}
\caption{\label{centered}
Total cross section for $\phi$ production in the reaction $pp \to pp{\phi}$ as a function of excess energy
$\sqrt{s}-\sqrt{s_{th}}$. For notation see the text.}
\end{figure}
\begin{figure}[h!]
\centerline{\epsfig{file=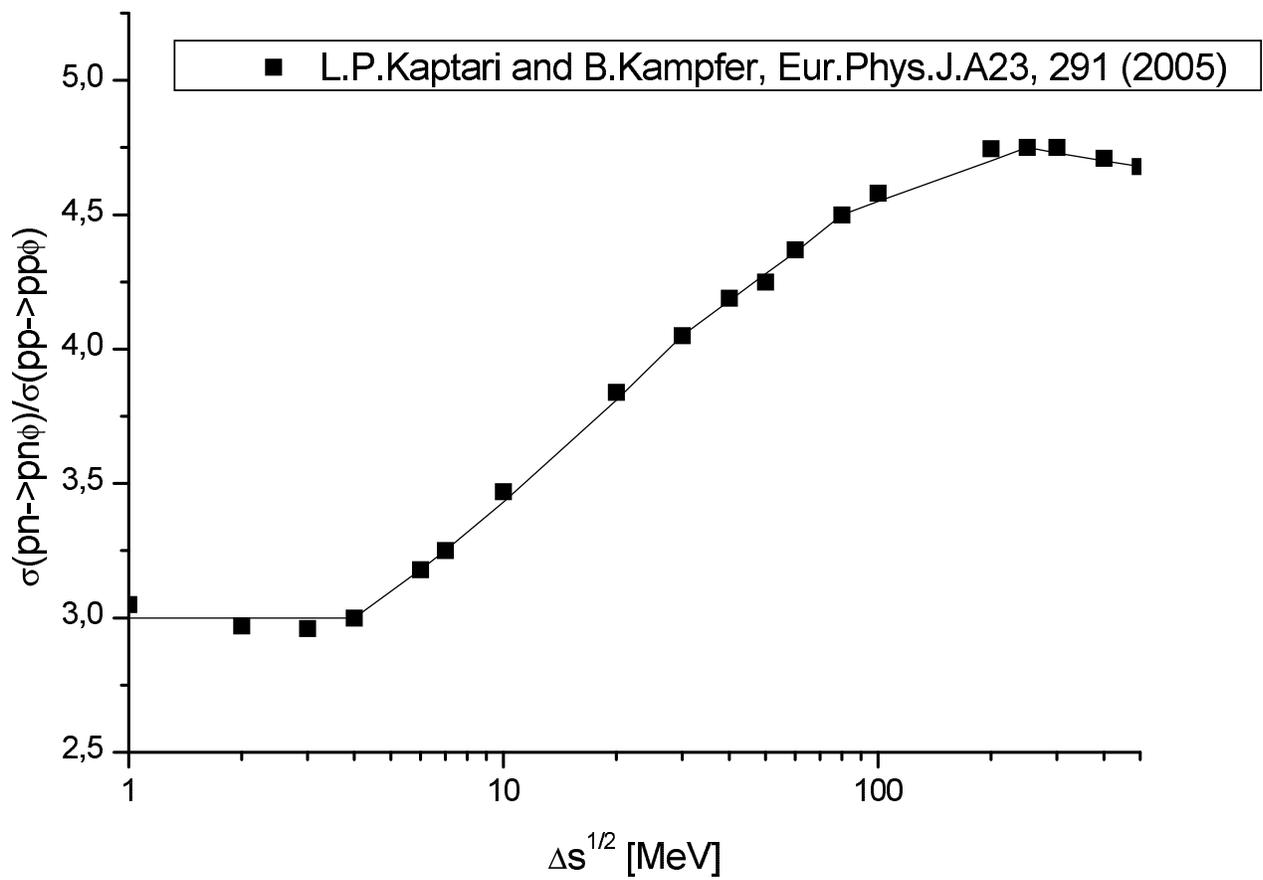,width=.88\textwidth,angle=270,silent=,
clip=}}
\caption{\label{centered}
Ratio of the total cross sections of $\phi$ meson 
production in $pp$ and $pn$ channels. For notation see the text.}
\end{figure}
\begin{figure}[h!]
\centerline{\epsfig{file=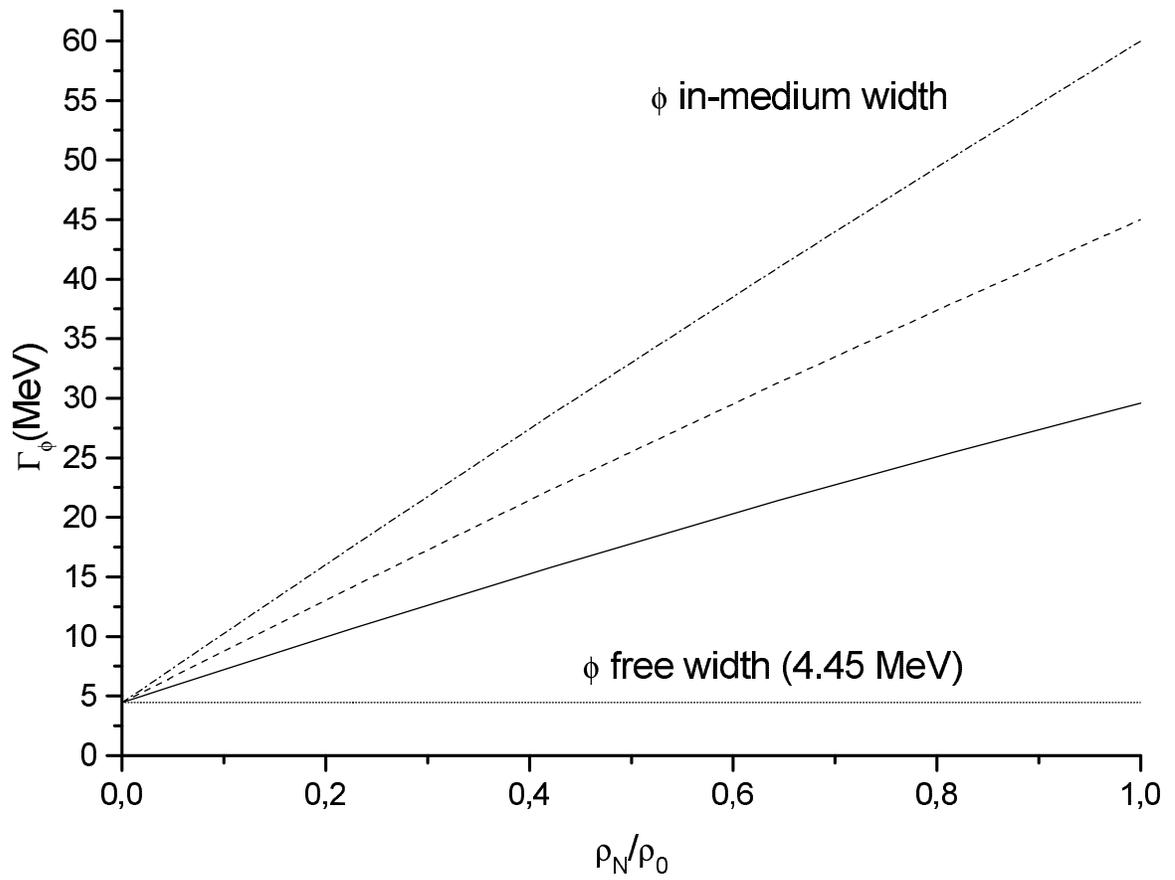,width=.88\textwidth,angle=270,silent=,
clip=}}
\caption{\label{centered}
Phi meson total width as a function of density. For notation see the 
text.}
\end{figure}
\begin{figure}[h!]
\centerline{\epsfig{file=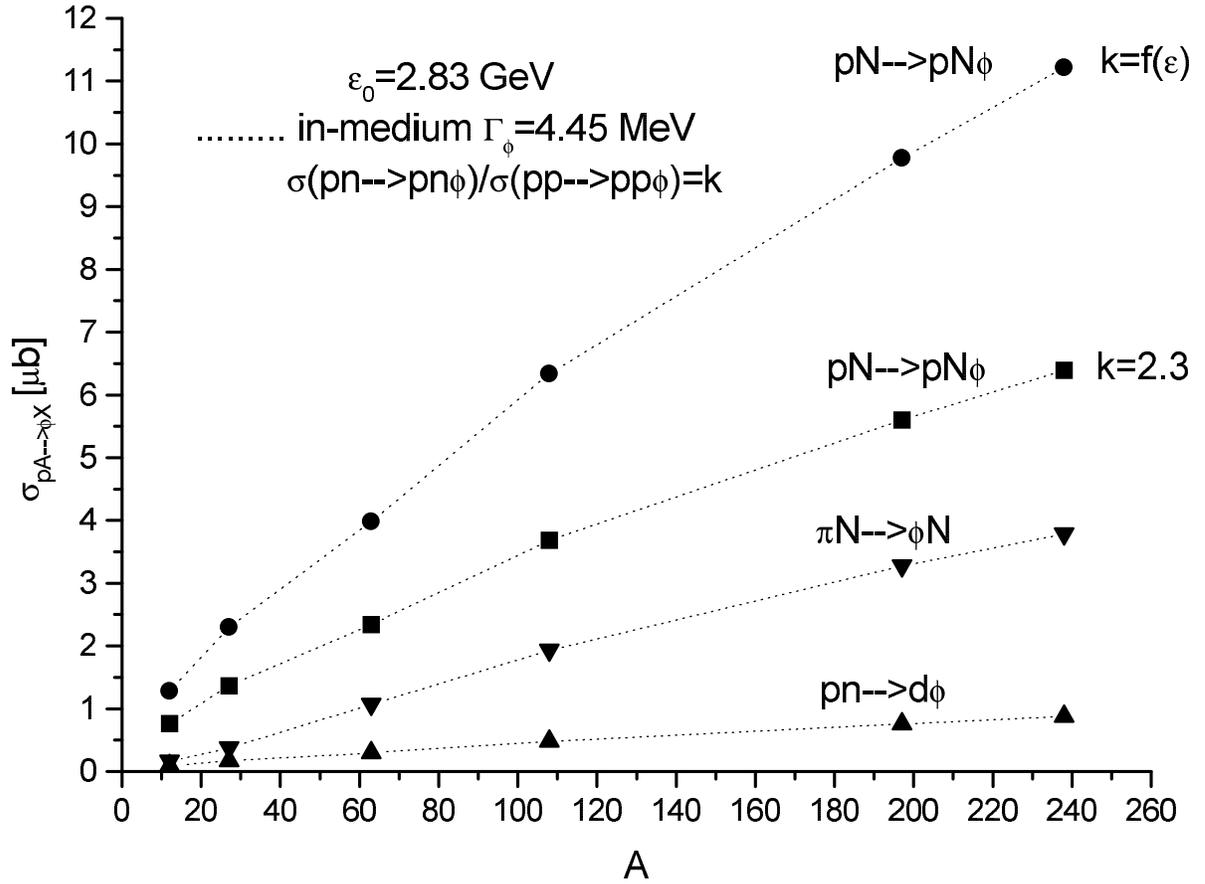,width=.88\textwidth,angle=270,silent=,
clip=}}
\caption{\label{centered}
A-dependences of the total cross sections of $\phi$ production by 2.83 GeV protons from primary
$pN \to pN{\phi}$ channels (1), (2) for two options (20), (21) for the cross section ratio   
$\sigma_{pn \to pn{\phi}}/{\sigma_{pp \to pp{\phi}}}$, from $pn \to d{\phi}$ process (3) as well as from secondary
 ${\pi}N \to {\phi}N$
channel (29) in the phi momentum range $0.6~{\rm {GeV/c}} \le p_{\phi} \le 1.8~{\rm {GeV/c}}$ and in the phi polar angle
domain $0^{\circ} \le \theta_{\phi} \le 8^{\circ}$ in the lab system--regions which are considered in the analysis of the
ANKE-at-COSY experiment [21, 22]. The absorption of phi mesons in nuclear matter was determined by their free width
(dotted curve in Fig. 3). The lines are included to guide the eyes.}
\end{figure}
\begin{figure}[h!]
\centerline{\epsfig{file=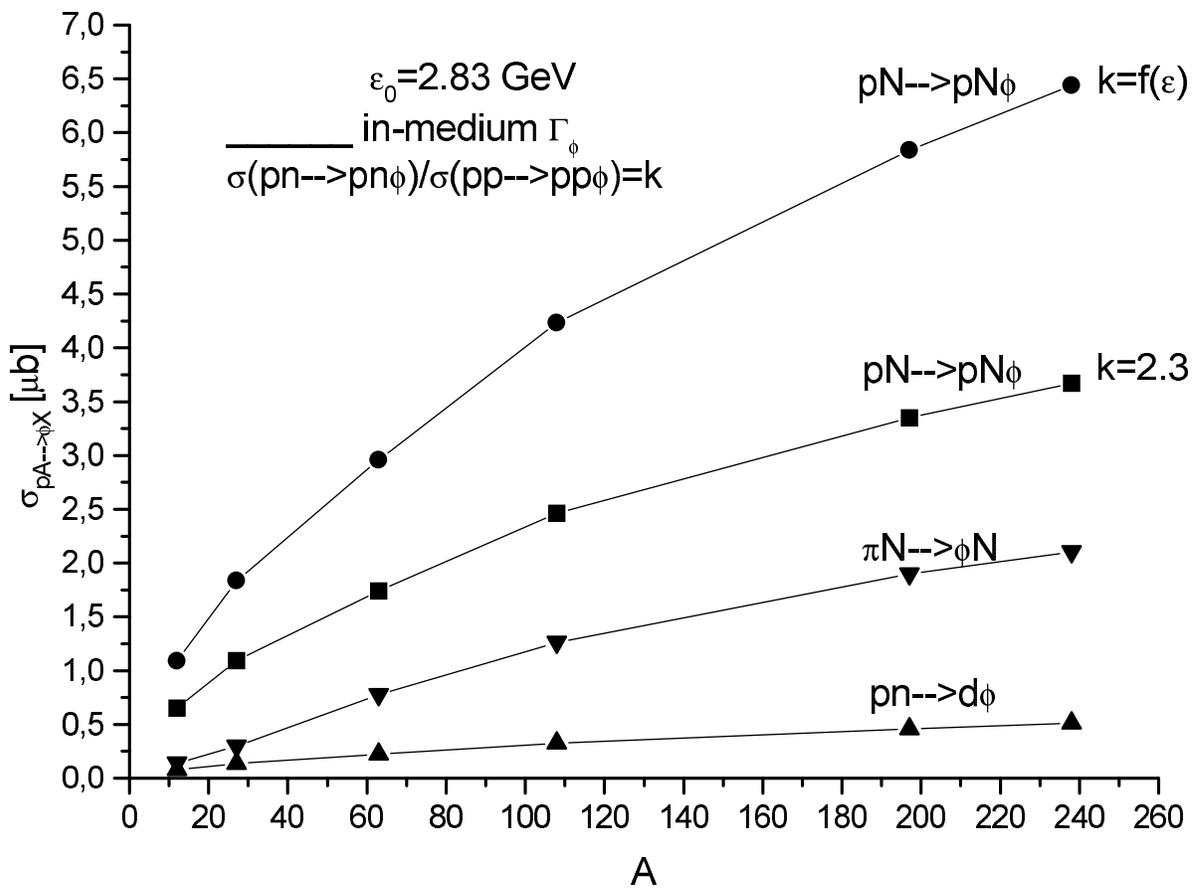,width=.88\textwidth,angle=270,silent=,
clip=}}
\caption{\label{centered}
The same as in Fig. 4, but it is supposed that the absorption of phi 
mesons in nuclear matter was governed by
their in-medium width shown by solid curve in Fig. 3.}
\end{figure}
\begin{figure}[h!]
\centerline{\epsfig{file=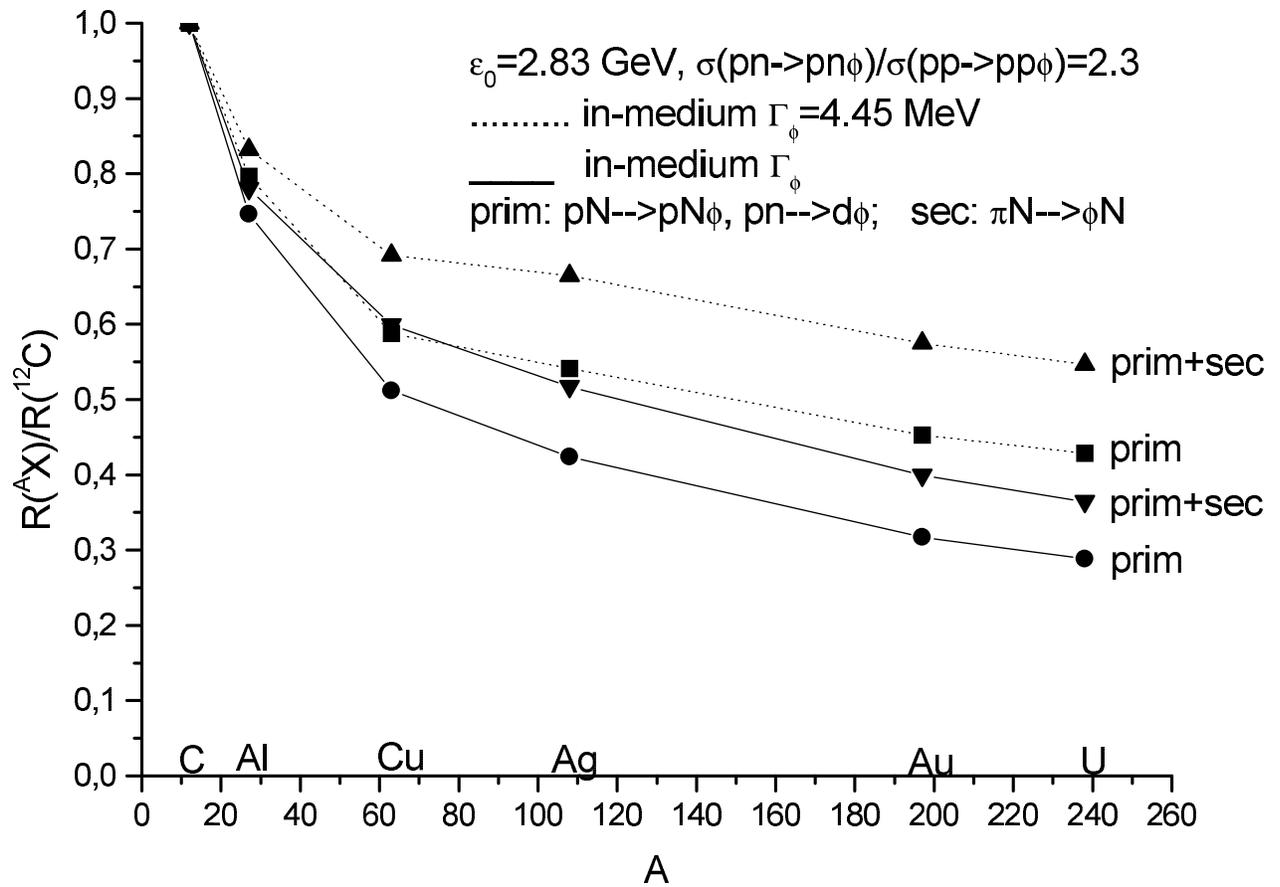,width=.88\textwidth,angle=270,silent=,
clip=}}
\caption{\label{centered}
Ratio $R(^{A}X)/R(^{12}{\rm C})$=
$({\tilde \sigma}_{pA \to {\phi}X}/A)/({\tilde \sigma}_{p^{12}{\rm C} \to {\phi}X}/12)$ as a function of the nuclear mass
number for initial energy of 2.83 GeV and for the cross section ratio $\sigma_{pn \to pn{\phi}}/{\sigma_{pp \to pp{\phi}}}=2.3$
calculated within the different scenarios for the $\phi$ meson production mechanism and for its in-medium width. For notation see
the text.}
\end{figure}
\begin{figure}[h!]
\centerline{\epsfig{file=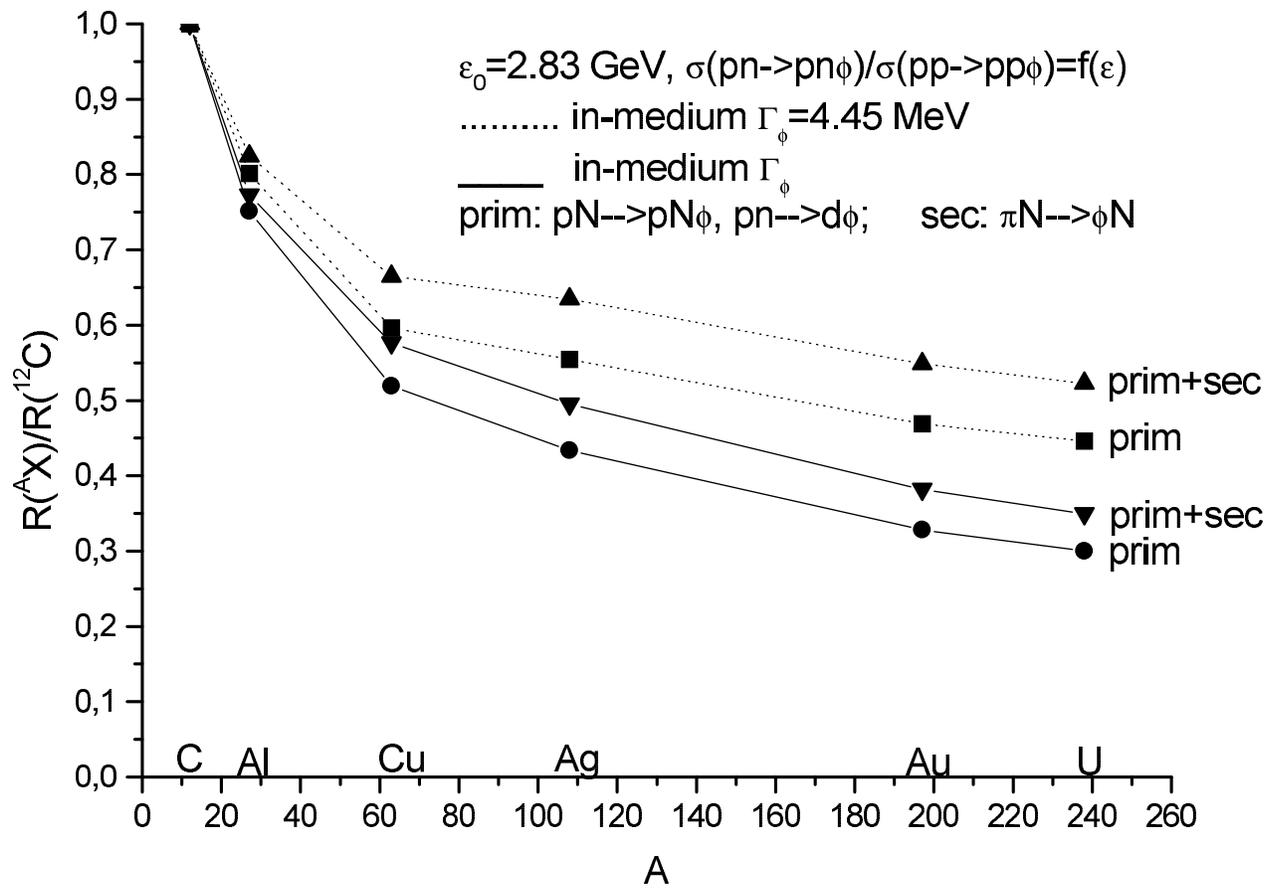,width=.88\textwidth,angle=270,silent=,
clip=}}
\caption{\label{centered}
The same as in Fig. 6, but calculated for the cross section ratio
$\sigma_{pn \to pn{\phi}}/{\sigma_{pp \to pp{\phi}}}$ in the 
excess-energy-dependent form (21).}
\end{figure}
\begin{figure}[h!]
\centerline{\epsfig{file=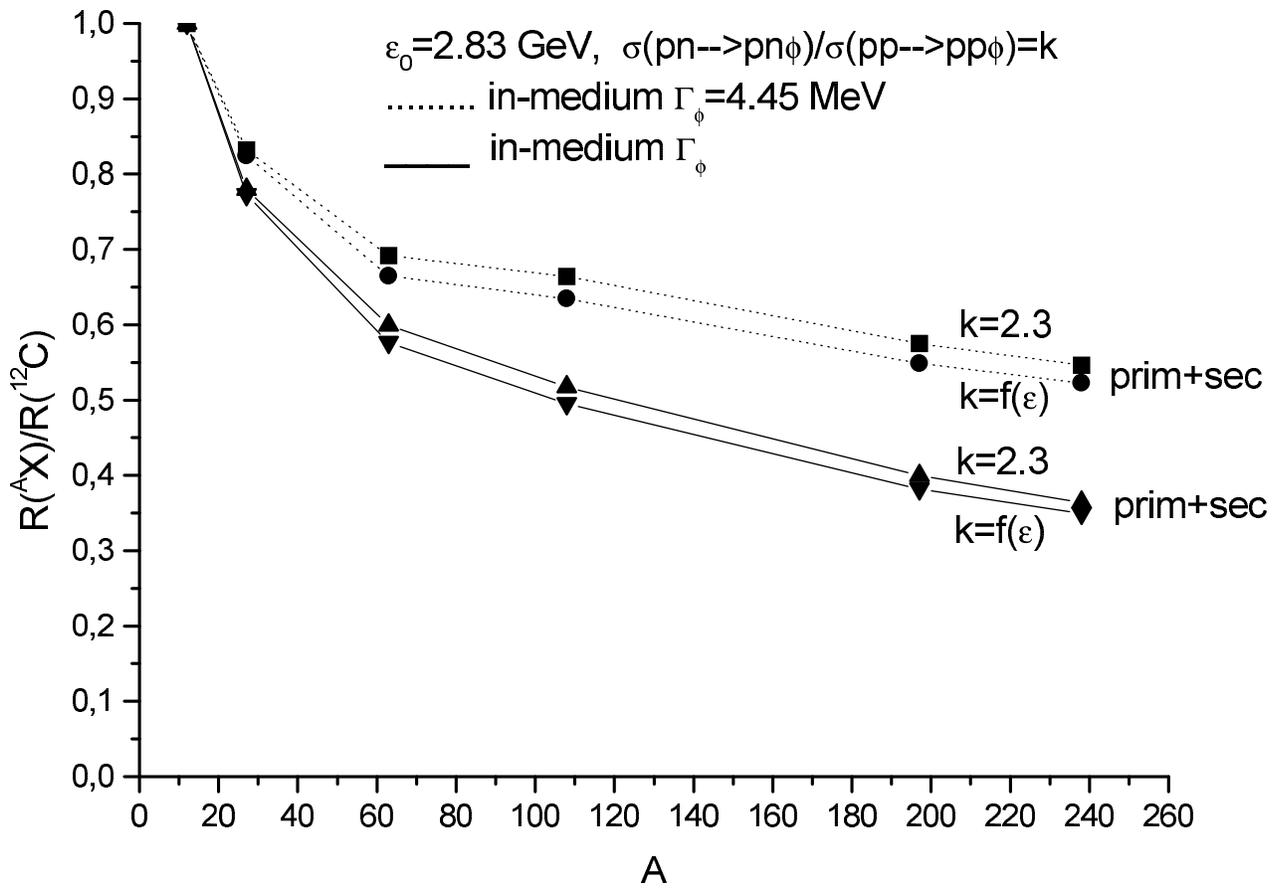,width=.88\textwidth,angle=270,silent=,
clip=}}
\caption{\label{centered}
Ratio $R(^{A}X)/R(^{12}{\rm C})$=
$({\tilde \sigma}_{pA \to {\phi}X}/A)/({\tilde \sigma}_{p^{12}{\rm C} \to {\phi}X}/12)$ as a function of the nuclear mass
number for the one- plus two-step $\phi$ production mechanisms calculated for incident energy of 2.83 GeV within the different
scenarios for the phi in-medium width and for the cross section ratio $\sigma_{pn \to pn{\phi}}/{\sigma_{pp \to pp{\phi}}}$.
For notation see the text.}
\end{figure}
\begin{figure}[h!]
\centerline{\epsfig{file=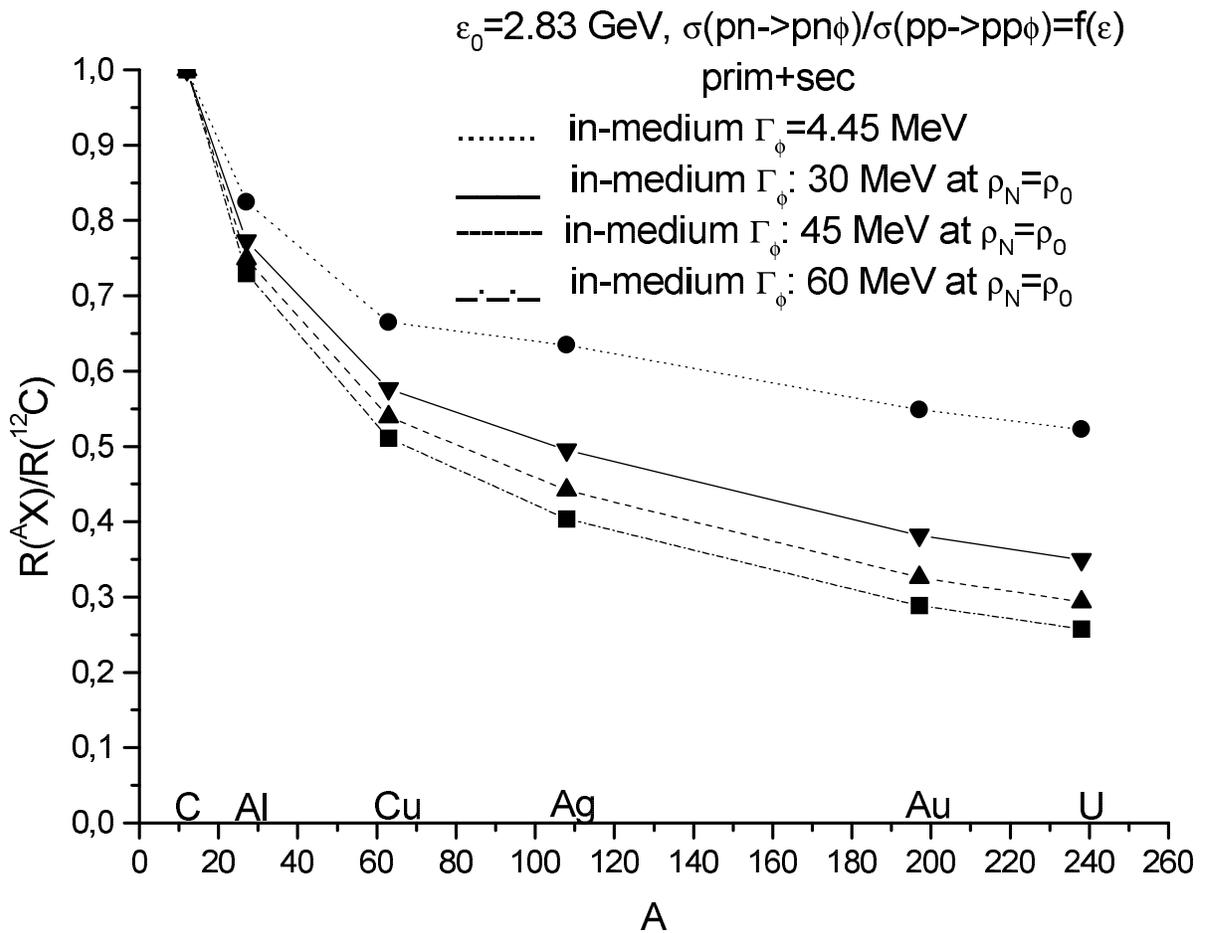,width=.88\textwidth,angle=270,silent=,
clip=}}
\caption{\label{centered}
Ratio $R(^{A}X)/R(^{12}{\rm C})$=
$({\tilde \sigma}_{pA \to {\phi}X}/A)/({\tilde \sigma}_{p^{12}{\rm C} \to {\phi}X}/12)$ as a function of the nuclear mass
number for the one- plus two-step $\phi$ production mechanisms calculated for incident energy of 2.83 GeV and for
the cross section ratio $\sigma_{pn \to pn{\phi}}/{\sigma_{pp \to pp{\phi}}}$ in the excess-energy-dependent form (21)
within the different scenarios for the phi in-medium width. 
For notation see the text.}
\end{figure}
\end{document}